\documentclass[a4paper, 11pt]{article}
\pdfoutput=1

\usepackage{jheppub}
\usepackage{amsmath}
\usepackage{graphicx}
\usepackage[titletoc,page]{appendix}
\usepackage{xspace}
\usepackage{subfig}
\usepackage{wrapfig}
\usepackage[outercaption]{sidecap}
\usepackage[labelfont=bf]{caption}
\usepackage{hyperref}
\hypersetup{colorlinks=true,
		    urlcolor=blue}

\newcommand{\pythia}{{\sc Pythia8}\xspace}
\newcommand{\madgraph}{{\sc MadGraph}\xspace}
\newcommand{\herwig}{{\sc Herwig++}\xspace}

\newcommand{\numpy}{{\sc Numpy}\xspace}
\newcommand{\scipy}{{\sc Scipy}\xspace}
\newcommand{\sklearn}{{\sc Scikit-Learn}\xspace}
\newcommand{\matplotlib}{{\sc Matplotlib}\xspace}
\newcommand{\rootcern}{{\sc ROOT}\xspace}

\newcommand{\pt}{\mbox{$p_{T}$}\xspace}
\newcommand{\pthat}{\mbox{$\hat{p_{T}}$}\xspace}

\title{Jet-Images:\\ Computer Vision Inspired Techniques for Jet Tagging}

\author[a]{Josh Cogan}
\author[a]{Michael Kagan}
\author[a]{Emanuel Strauss}
\author[a]{Ariel Schwartzman}

\affiliation[a]{SLAC National Accelerator Laboratory\\Menlo Park, CA 94028, USA}

\emailAdd{joshgc@slac.stanford.edu}
\emailAdd{makagan@slac.stanford.edu}
\emailAdd{estrauss@slac.stanford.edu}
\emailAdd{sch@slac.stanford.edu}

\abstract{We introduce a novel approach to jet tagging and
  classification through the use of techniques inspired by computer
  vision. Drawing parallels to the problem of facial recognition in
  images, we define a \textit{jet-image} using calorimeter towers as
  the elements of the image and establish jet-image preprocessing
  methods.  For the jet-image processing step, we develop a discriminant for classifying the jet-images
  derived using Fisher discriminant analysis. The effectiveness of the
  technique is shown within the context of identifying boosted
  hadronic $W$ boson decays with respect to a background of quark- and
  gluon-initiated jets. Using Monte Carlo simulation, we demonstrate that the performance of this
  technique introduces additional discriminating power over other substructure approaches, 
  and gives significant insight into the internal structure of jets.
  }

\begin{document} 
\maketitle
\flushbottom

\section{Introduction}
\label{sec:intro}

Modern TeV scale colliders, such as the LHC~\cite{LHCMachine}, are capable of producing highly boosted jets from the decay of heavy particles.  The large momentum and mass of the heavy particle may lead to a boosted system in which the heavy particle's decay products merge into a single jet. In order to discover and study such systems, it is vital to discriminate boosted heavy particle jets from those of  QCD initiated quark/gluon
jets (which we will refer to as QCD jets for the remainder of the text).  The techniques of jet substructure, which exploit the internal structure of the jet, have been developed for this task.  Contemporary experimental and theoretical work in the field of jet substructure includes, but is not limited to,  subjet finding and jet tagging / jet flavor identification algorithms~\cite{SeymourSearch,Plehn2012Top,BDRS2008,Kaplan2008Top,Thaler2008BoostTop,Thaler2011Nsub,Thaler2011Nsub2,Gall2008HiggsMVA,Gall2011QVG,Cui2012Wtag,Plehn2009Higgs}, jet grooming and pileup mitigation~\cite{Krohn2010Trim,Cacc013Area,EllisPrune}, template and matrix element techniques~\cite{Backovic2013HiggsTemplate,Soper2011Decon}, energy flow and event shapes~\cite{Hook2012Dipol,Larkoski2013ECorr,Ellis2010Evtshape,Almeida2009Evtshape,Berger2003Evtshape,Gall2012Color,Jank2012Anglular}, and probabilistic jet clustering~\cite{Kahawala2013JetSampling,Ellis2012QJets};  more inclusive summaries can be found in references~\cite{BOOST2012proceed,SubStrRev1,SubStrRev2,SubStrRev3}

Typical approaches to jet flavor identification and jet tagging begin with a simplified
analytic expression of the system's behavior, such as the two-pronged
nature of the hadronic $W$ boson decay, or the large particle multiplicity of
gluon initiated jets relative to quark initiated jets.  A variable which capitalizes on this analytic knowledge
and is ideally not subject to large theoretical
uncertainties is derived.  In contrast, our method
does not rely on an analytic description of the system, such
as colorflow information, but instead uses a catalog of jets produced by Monte Carlo simulation as a training sample to learn the discriminating features between several categories of jets.
The method begins by putting jets through a series of preprocessing steps to build a consistent representation of the jets.  After preprocessing, we employ a fast, 
linear, and powerful method for feature extraction and physical interpretation\footnote{ Sophisticated
non-linear multivariate techniques are also an option but often suffer from long
training and testing times, and provide less physical
intuition to the underlying discriminating features}.

Our methods for preprocessing and discrimination are inspired by techniques in the field of computer vision.  Most notably, our task is similar to that of facial recognition, and we have developed jet-specific analogs to the algorithms used in facial recognition. As such, the work presented here serves to both build a connection between the fields of jet substructure and computer vision, and as an introduction to the use of computer vision techniques in the analysis of hadronic final states.  The basic algorithm presented here is thus serving as a general proof of principle, and can be more deeply developed in future studies.

In Section~\ref{sec:algorithm} we present the``jet-image" algorithm, which both describes the definition of the jet-image, the jet-image preprocessing steps, and the jet-image processing (that is, the methods of discrimination).  Section~\ref{sec:samples} describes the samples used for our performance studies of the algorithm, and Section~\ref{sec:studies} presents the case study  where the algorithm is employed for discriminating hadronically decaying W jets from QCD jets and is compared to other jet substructure techniques.  In Section~\ref{sec:conclusion} we discuss the information that can be gained from our approach, and further possible application and developments.
\section{Algorithm}
\label{sec:algorithm}

All operations
described bellow are performed in Python (v2.7.2) using the
\numpy~\cite{numpy} and \scipy~\cite{scipy} libraries.  In addition, many of the
 classifiers are built using \sklearn~\cite{scikit-learn}, and the
figures are made using \matplotlib~\cite{Hunter:2007} and \rootcern~\cite{ROOTref}.

\subsection{Jets as Images}
\label{sec:image}

For this study, we restrict the inputs to jet reconstruction strictly to calorimeter towers
and, with little loss in generality, approximate the calorimeter as a
single layered grid of towers\footnote{As practically all the
particles in a jet deposit their energy in the calorimeters, the
idealized calorimeter object sums and stores the energy of all the
particles entering a given tower, excluding neutrinos and muons.} 
with spacing $\Delta \eta \times \Delta
\phi=0.1 \times 0.1$ spanning $[-2.5,2.5]$\footnote{This $\eta$ coverage was chosen as it roughly corresponds to the  region which is most commonly used for physics analysis by the ATLAS~\cite{Aad:2008zzm} and CMS~\cite{CMSDesign} experiments as it is covered by both calorimeter and tracking detectors.} in $\eta$ and
$[0,2\pi]$ in $\phi$.  Using common jet-finding algorithms~\cite{KTAlgo1, KTAlgo2, CAAlgo, AKTAlgo, Shelton2013TASI}, we
identify the angular location of the jets (as the jet axis) and
save all towers within a $2R$ by $2R$ square, centered on that
location for further examination.   This yields the $(2\frac{R}{0.1}+1)\times
(2\frac{R}{0.1}+1)$ grid\footnote{The additional tower ensures that if
  the jet axis is at the center of a tower, all calorimeter towers
  within a radius of $R$ are contained in the grid} of towers which we
denote the \textit{jet-image}, an example of which can be seen in
Figure~\ref{subfig:orig}. Henceforth, we will refer to calorimeter towers which are included in a jet-image as pixels.
It is important to note that with this jet-image
representation the similarity of two jets can be established using the
dot product between two jets.  That is, an $N\times N$ pixel jet-image
$A$ can be recast into a $N^{2}$ dimensional vector $\bar{A}$, where the elements $A_{i,j} = \bar{A}_{i+N\times (j-1)}$ with $i,j \in \{1...N\}$, and the dot product between jet-images $A$ and $B$ is 
\begin{equation}
A\cdot B = \sum_{k=1}^{N^{2}} \bar{A}_{k}\bar{B}_{k}
\end{equation}

While this study focuses on jet-images built from calorimeter towers, any inputs which can be
mapped to a pixelated image could be used to build jet images.  For instance, jet-images could be build from
tracks, calorimeter clusters, truth particles, etc., all of which could in principle use much finer pixel granularity due to the higher resolution of the inputs.  Once the jet-images are built from the inputs, the algorithms described in the proceeding text are directly applicable.

Jet-images have several properties that make them useful inputs to
jet flavor tagging techniques.  First, each image has the same
dimensionality---every jet can be described by a fixed set of
numbers. This is an important property for the application of image
processing techniques and for training classification algorithms. 
Second, as a low-level jet description, 
the jet-image uses all available
information for later discriminatory techniques rather than compressing the information into a set of derived variables. Currently, the information used per pixel is  the total deposited transverse energy but could be extended to include further information, such as the longitudinal segmentation of energy deposits in the calorimeter.  However, we leave these extension for future studies. Third, the similarity between two jets can be easily and quickly computed using standard linear
algebra computations.  While the concept of analyzing jets as images is not new (see for instance~\cite{FFTjet,TaitWavelet,DreminWavelet,WaveletMonk}), this paper presents a computer vision inspired framework for processing and interpretation of jet-images.

Having re-cast the representation of the calorimeter information, we
introduce the jet-image preprocessing steps and the discrimination
techniques used for flavor identification.

\subsection{Jet-Image Preprocessing}
\label{sec:preprocessing}
Approaches to facial recognition (see~\cite{Zhao2003FaceRec} and
references therein for review) attempt to classify a face in an image
by learning the structure of the face, i.e. learning the expected
distribution of pixel intensities from a set of example images. The
ability to accurately recognize a face in an image can be sensitive to
lighting conditions, intensity, shadows, the orientation of the
faces, facial expressions, noise, etc. These features 
introduce large variations in the pixel intensities but are not
correlated with different categories and thus obfuscate
the underlying discriminatory distributions.
Image preprocessing steps are used to standardize the images
and greatly improve later classification performance.

Analogously, our approach to jet tagging attempts to learn the structure of 
the transverse energy distribution of the pixels of the jet-image. The use of
preprocessing is paramount to accurate jet-image classification,
and as such we have developed a set of preprocessing
steps specifically for this context.  Many of these steps have parallels 
in facial recognition, which should become evident from the description.
The jet-image preprocessing proceeds as follows:

\begin{enumerate}
\item \textbf{Noise Reduction}: The effects of noise in the jet-image
  from pileup are reduced using trimming~\cite{Krohn2010Trim}.  Trimming is used due to its simplicity, the fact that it uses subjets which are used a later stages in the algorithm, and because it is seen to
  greatly reduce the impact of pileup on the algorithm performance.  Different jet grooming techniques (see~\cite{SubStrRev2,Shelton2013TASI} and references therein) which mitigate the effects of pileup could also be used.

\item \textbf{Point of Interest Finding}: The positions of the leading
  regions of transverse energy deposition are located using the subjets  resulting from the trimming procedure.  Subjets are ordered by $p_{T}$ to decide the relevance of the points of interest.  The number of subjets required
  depends on the alignment procedure (step 3), but at least one
  subjet must be identified. This step, in the case of a 2-prong decay such as a W-jet, is similar to locating the
  eyes in an image of a face.

\item \textbf{Alignment}: The jet-image is aligned such that the
  relative pixel location of primary features are always the same.
  This step exploits symmetries of the $\eta-\phi$ space and
  removes variations in jets which do not improve classification
  accuracy.  This is analogous to aligning the eyes of a face within
  an image to always lie in the same pixels.  There are three
  components to jet-image alignment:
  \begin{itemize}
  \item \textit{Rotation}: Rotation is performed to remove the
    stochastic nature of the decay angle relative to the $\eta-\phi$
    coordinate system.  This alignment can be done very generally, by
    determining the principle axis~\cite{Sylvester1852PincpAxes} of the original image and rotating
    the imagine around the jet-energy centroid such that the principle
    axis is always vertical. Alternatively, process specific
    information can be used. For two-body decay processes (such as the
    hadronic decay of a W boson) the direction connecting the axes of the
    leading two subjets can be rotated until the leading
    subject is directly above the subleading subjet.  However using
    process specific information can lead to acceptance loss, 
    for example when the two decay products merge into a single subjet.

  \item \textit{Translation}: Once rotated, the jet-image is translated
    such that the jet-image energy centroid or leading subjet is always centered in the same
    pixel. This procedure is paramount to adjusting the position of the eyes
    in a picture, anchored so as to always appear in the same pixels.

  \item \textit{Reflection}: Once translated, the image is reflected
    over the vertical axis such that the side of the image with
    maximal transverse energy always appears on the right side of the image.  This ensures that
    the hardest radiation always appears in similar locations and
    can be exploited by the training procedure for classification.

\end{itemize}

\item \textbf{Equalization}: Jet-images are normalized such that the
  dot product of a jet-image with itself (i.e. the sum of squared pixel contents)
   is equal to one.  This removes the absolute energy scale
  dependence of jet-images, thus allowing for comparisons of
  jet-images with different energies.  This step is analogous to the
  standardizing the lighting conditions of images.

\item \textbf{Binning}: In many cases, the expected jet-images may
  vary significantly with a known variable;
  in this case, the variable can divide a
  class of jet-images into a set of sub-classes with more 
  uniform jet-images.  For instance, if the total transverse energy of the
  jet-image or the $\Delta R$ between the subjets causes significant
  variations, jet images can be binned into different ranges
  of the variable.  This is analogous to separating images based
  on the facial expression.
  A different discriminant can then be trained
  separately for each sub-class.

\end{enumerate}

\begin{figure*}[htb]
  \centering
  \subfloat[Jet-image prior to rotation\label{subfig:orig}]{
    \includegraphics[width=0.3\textwidth]{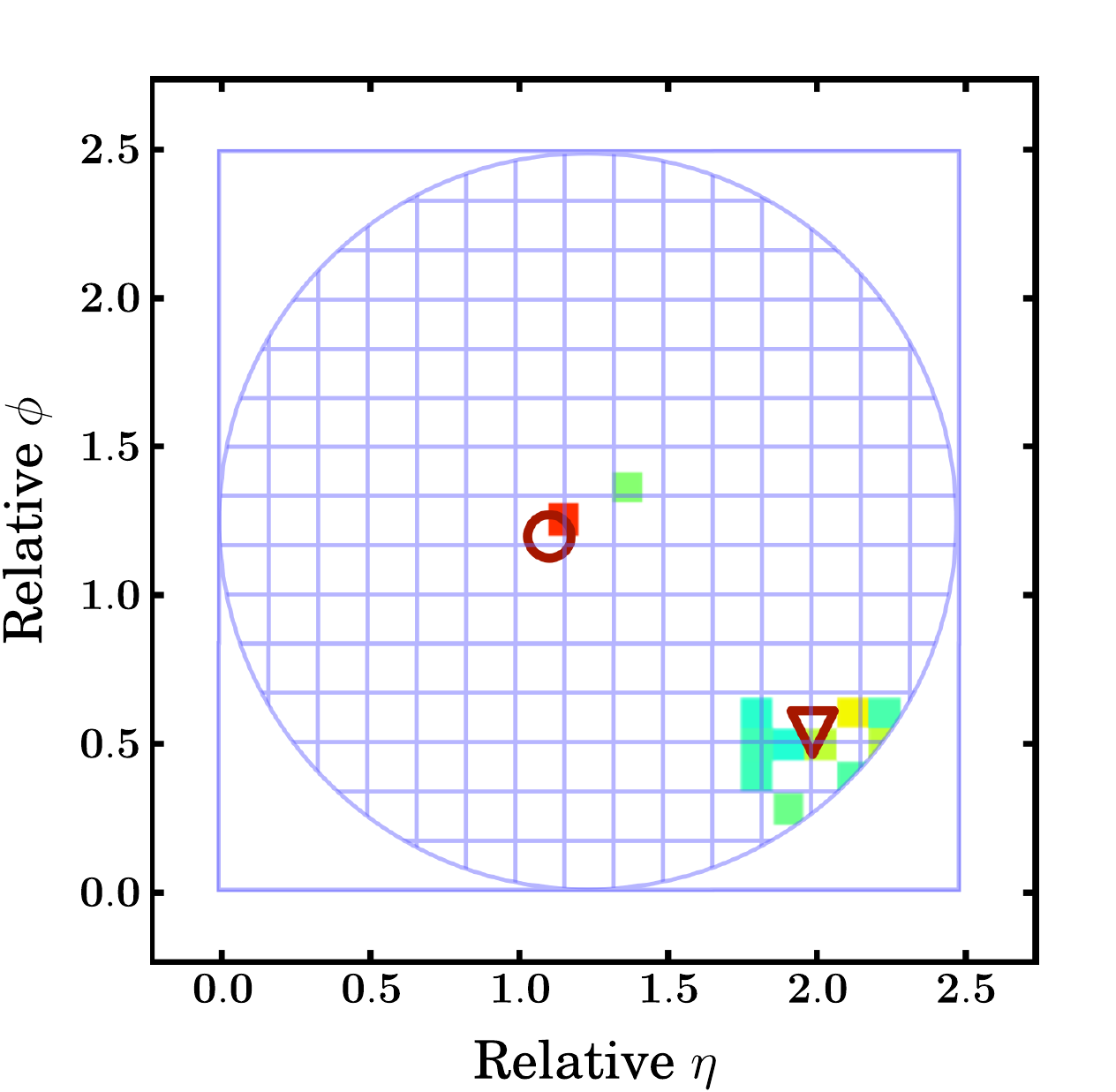}
  }
  \hfill
  \subfloat[Rotated pixel grid\label{subfig:rot_grid}]{
    \includegraphics[width=0.3\textwidth]{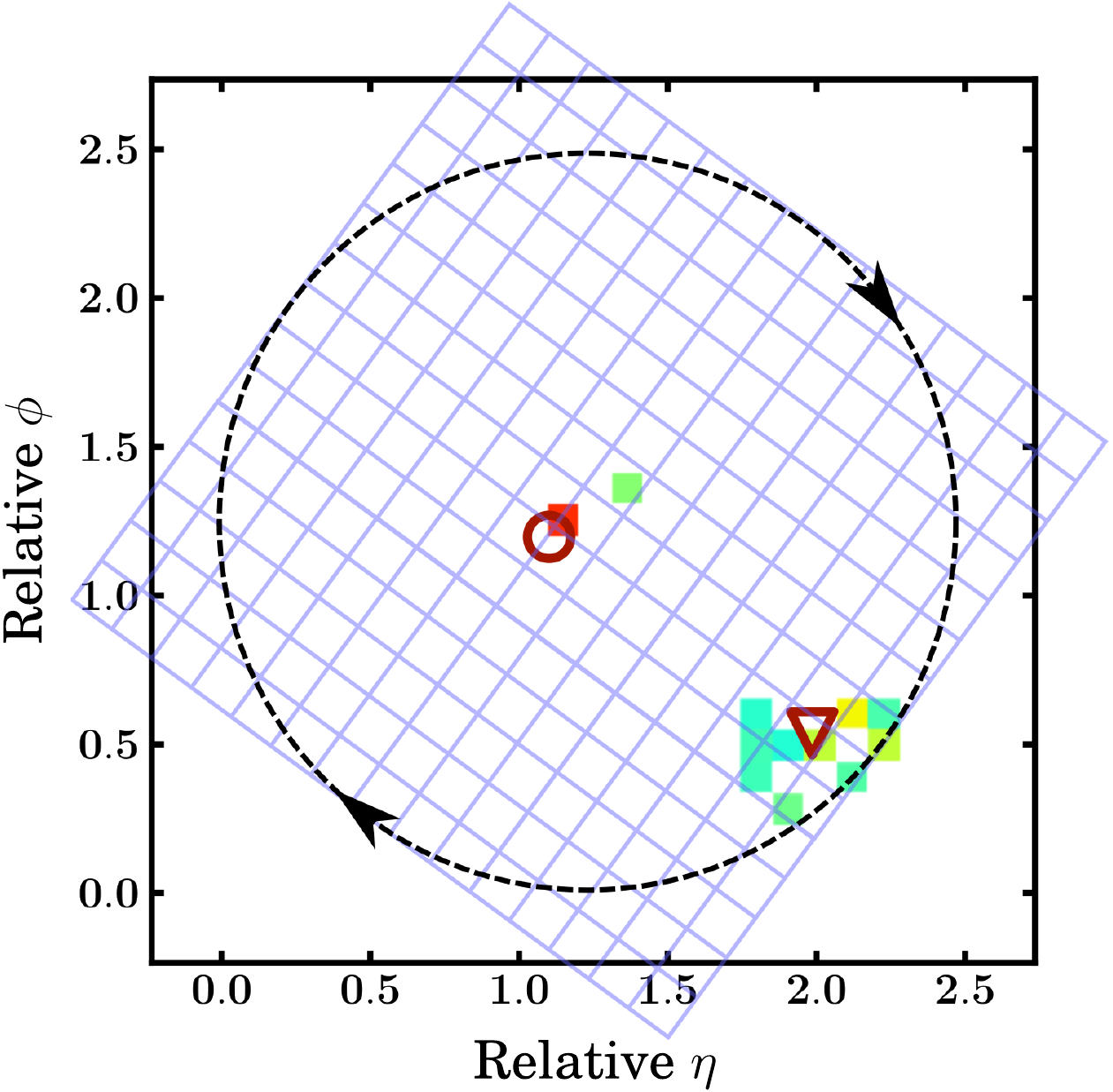}
  }
  \hfill
  \subfloat[Jet-image after projection onto rotated grid, before translation\label{subfig:proj}]{
    \includegraphics[width=0.3\textwidth]{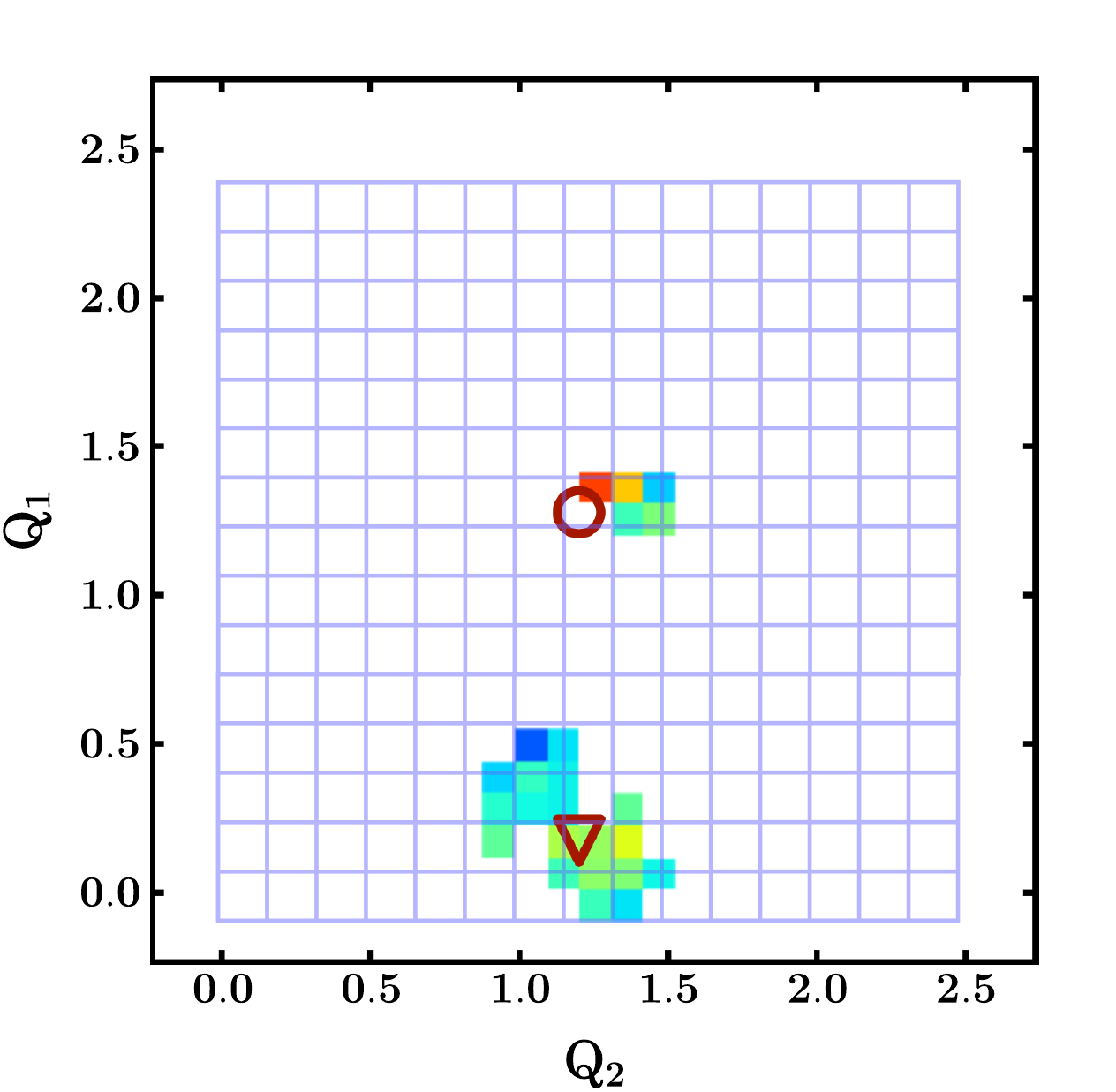}
  }
  \\
  \subfloat[Average jet-image, prior to rotation\label{subfig:orig_avg}]{
    \includegraphics[width=0.3\textwidth]{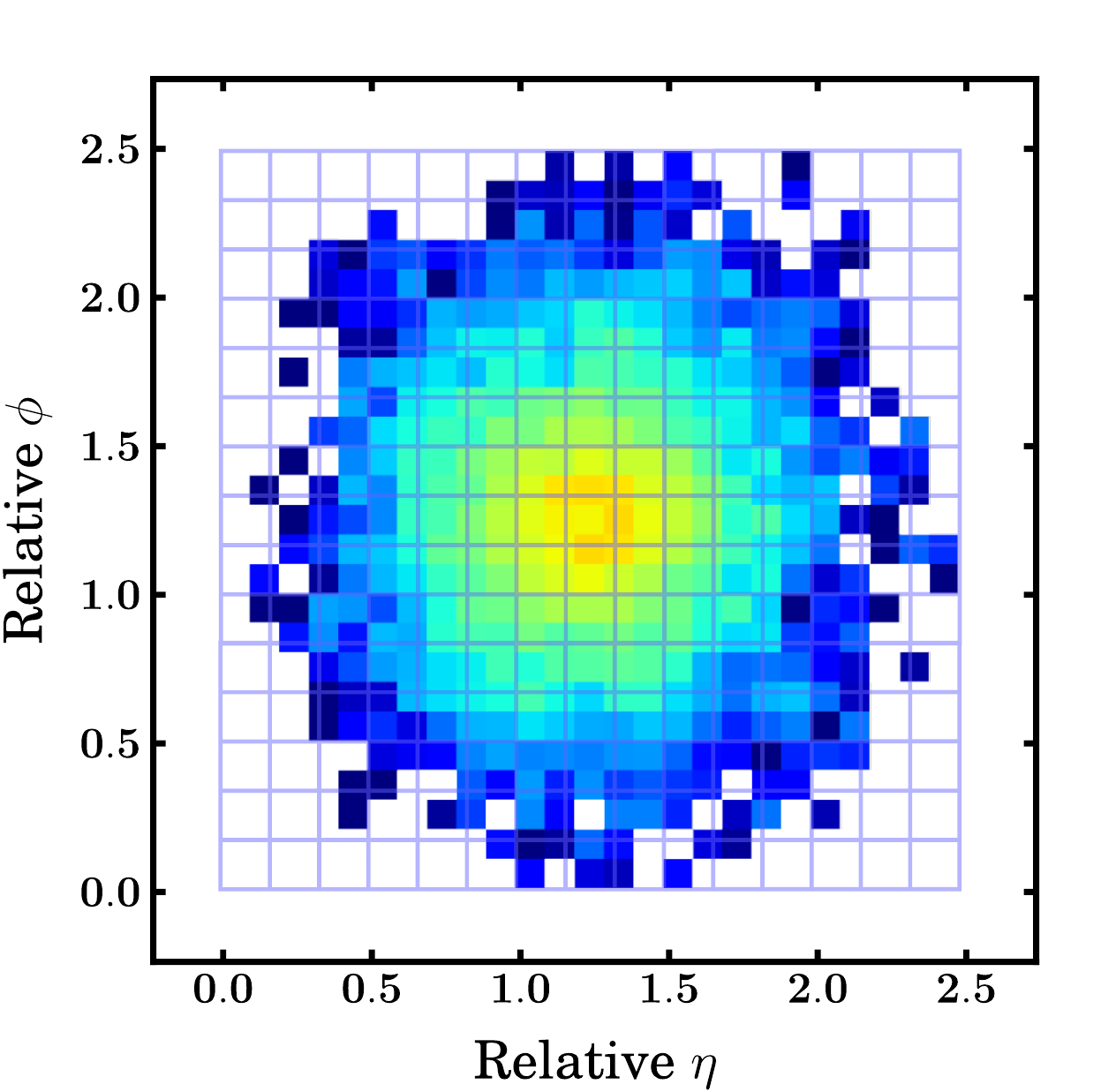}
  }
  \hfill
  \subfloat[Average jet-image, after preprocessing\label{subfig:proj_avg}]{
    \includegraphics[width=0.3\textwidth]{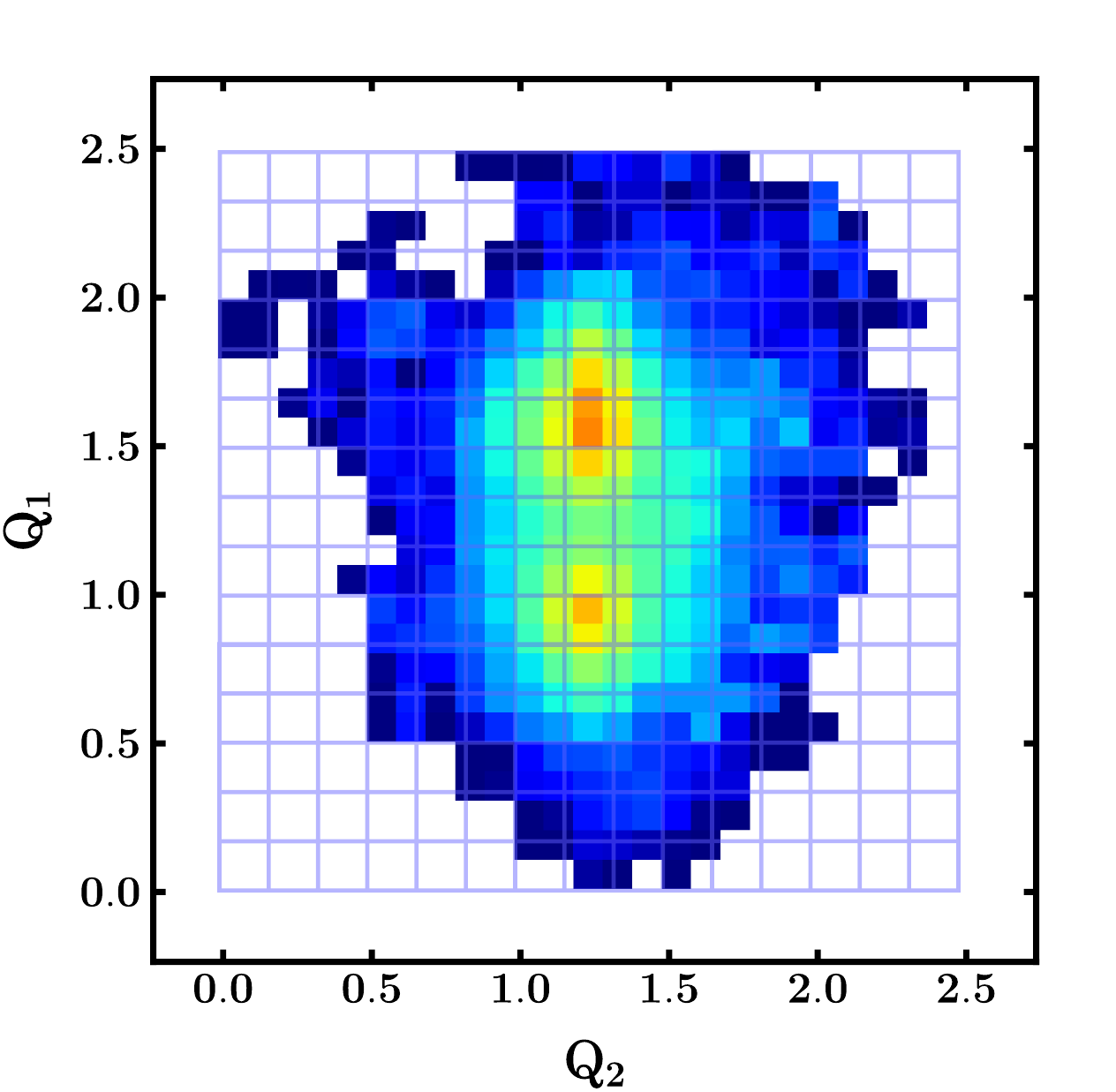}
  }
  \caption{The preprocessing of jet-images and the impact on the average jet-image 
  for W jets in which the leading jet with \pt between $200$ and
  $250$~GeV. Note that the grid in figure~\ref{subfig:proj} appears shifted down to represent the jet-image before  translation, which is subsequently translated such that the leading subjet lies in the location ($Q_{1}\sim1.5, Q_{2}\sim1.25$) as see in the final average jet-image of figure~\ref{subfig:proj_avg}.\label{fig:rot_jets}}
\end{figure*}

An example of the image preprocessing with jets from hadronically
decaying W bosons can be seen in Figure~\ref{fig:rot_jets} plotted using the $\eta$ and $\phi$ coordinates of the pixels relative to the jet axis (before rotation) and using the rotated coordinate system $Q_{1}$ and $Q_{2}$ after rotation.  The
average of a large sample of W jet-images without preprocessing can
be seen in Figure~\ref{subfig:orig_avg}, where there is no clear sign
of the two-prong decay structure of the W jet.  Figures~\ref{subfig:orig},
~\ref{subfig:rot_grid}, and ~\ref{subfig:proj} illustrate the
different stages of preprocessing on a single W jet-image.  Finally,
Figure~\ref{subfig:proj_avg} shows the average of the same large sample
of W jet-images after preprocessing with the two-prong 
structure of the W decay clearly visible.

This preprocessing prescription is generic enough to be applied
to most jet flavor identification problems. However,
the impact of the preprocessing steps on
the system under consideration must be carefully considered and
in some cases tailored. 
For example, while trimming, or more generally jet grooming, is important for LHC experiments which expect
large amounts of pileup, it could be dropped for use in a cleaner 
experimental environment. Similarly, the rotation steps can be adjusted depending on the 
system under study (i.e. two-prong decays or three-prong decays).  

\subsection{Jet-Image Processing: Constructing the Discriminant}
\label{sec:fisherjet}

Having preprocessed the jet images, we have a representation which can
be thought of as a vector specifying a coordinate in a high-dimensional space. 
Linear classification of the jets is performed by
projecting the vector onto the discriminate direction which ideally
maximizes the separation between different classes of jets.

A number of linear classifiers can be used for discrimination. In facial recognition,
Principle Components Analysis (PCA)~\cite{Pearson1901PCA} and Fisher's Linear Discriminant
(FLD)~\cite{Fisher:1936} are common choices and their specified directions are known as
\textit{eigenfaces} and \textit{fisherfaces}
\cite{Belhumeur96eigenfacesvs} respectively. 
While both methods have been investigated within the context jet-images, FLD is observed to perform significantly better and thus
this paper focuses on FLD exclusively.
FLD identifies the plane in the high dimensional feature space which
maximizes the separation between the jet classes and simultaneously minimizes the scatter within each jet class
\footnote{In general, FLD can be trained on M classes, with $M\geq2$, resulting in $M-1$ Fisher discriminants for separating the $M$ classes}.
Since FLD uses knowledge of the within-class variations,
it is less influenced by variations present in both classes than PCA. In addition, we employ a regularized FLD implementation~\cite{Zhang2010Disc} which mitigates the impact of statistical fluctuations in the training sample and reduces overfitting.

FLD is trained using a set of preprocessed example jet-images from two classes 
and produces a discriminant $F$, which we denote the \textit{Fisher-jet}, that has the same dimensionality as the
example jet-images and thus can be viewed as a jet-image itself. Discrimination
between classes for a jet-image, $A$, is then achieved by projecting $A$ onto the Fisher-jet. That is,
\begin{equation}
D[A] = \sum_{k=1}^{N^{2}} \bar{F}_{k}\cdot \bar{A}_{k}
\end{equation}
where $D[A]$ is the discriminant output for the jet-image $A$, and $\bar{F}$ and $\bar{A}$ are
the vector representations of $F$ and $A$, respectively, as discussed in section~\ref{sec:image}.

FLD produces a robust classifier which is fast to apply, fast to derive,
and can itself be interpreted as a type of
image. Therein lies much of the technique's power. For any system
under study, the discriminant reduces the dimensionality of the
problem, such that the properties of a jet can easily be expressed
numerically and a cut on the discriminant value $D$ can be used to
separate classes.  Since the discriminant output is computed by
performing a projection of a jet-image onto the Fisher-jet, positive values in the
Fisher-jet identify features which are indicative of a jet of
class A whereas class B is highlighted by negative coefficients. The magnitude of the coefficient indicates the
classification strength of the feature. To understand the behavior of the discriminant, the Fisher-jet
image reveals what differences are being exploited in a simple and
intuitive way.  This is an important aspect of the
technique. Such a visualization can be used to convince ourselves that
the algorithm is detecting relevant differences between the two
classes. Furthermore, the ability to visualize the classifier means
that this approach is also useful as a step in exploratory analysis,
and could be used to guide the construction of top-down analytical
solutions to a problem.  This aspect of our technique contrasts with
widely used multivariate techniques in high-energy physics, 
such as neural networks and boosted decision trees, which predominantly use
variables derived from the jet's kinematics and substructure rather than directly using
the jet constituents.

\section{Samples}
\label{sec:samples}

Monte Carlo samples are produced with \pythia
(v8.176)~\cite{Pythia8}, \madgraph (v1.5.11)~\cite{Madgraph5}, and
\herwig (v2.6.3)~\cite{HerwigPP}, with proton-proton collisions at
$\sqrt{s} = 8$~TeV. In all cases, the effect of pileup from multiple
collisions in one crossing of the collider beams is simulated by
adding the energy of particles from additional \pythia ``minBias'' interactions.  
The number of additional pileup interactions per bunch crossing is a random variable 
sampled from a Poisson distribution with mean $\mu$.  In this paper we consider
the cases of no pileup and  $\mu=30$.

The signal sample used for these studies is di-boson $WW$
production, with one $W$ decaying to a muon and neutrino, and one $W$
decaying to a pair of quarks.  The matrix element and shower for the boosted $W$ samples are
calculated with \pythia. The boost of the system is controlled by
applying a loose cut on the matrix-element \pthat, followed by a
harder cut on the leading reconstructed jet \pt such that it falls
within a 50~GeV-wide \pt window (e.g. $200 < \pt < 250$~GeV). The
minimum and maximum \pthat was set 50~GeV below and above the jet \pt
interval being studied. The corresponding light and gluon jet background is
simulated using \pythia with $W$ + jets, where the $W$ decays to a
muon and a neutrino. Again, a loose cut on \pthat is used to set a
generator-level boost for the system, followed by a harder cut on the
leading reconstructed jet. Similar samples are generated with \herwig
for validation studies.

Jet finding is performed with Fastjet
(v3.03)~\cite{fastjet} on the massless four-vectors defined by the calorimeter
towers.  For each generated event,  the towers are clustered using the
Cambridge/Aachen algorithm~\cite{CAAlgo} with a radius $R = 1.2$ and 
only jets with $p_{T} >25$~GeV are kept.  We apply trimming with
$k_T$ $R =0.3$ subjets and $f_{cut}=5\%$ to
reduce the impact of pileup interactions. 
This serves as the noise reduction step introduced in
Section~\ref{sec:preprocessing}, and no additional pileup subtraction was performed.

\section{Case Study: W Boson Jet Tagging}
\label{sec:studies}
In this section we demonstrate the performance of this technique in
classifying boosted $W \rightarrow q q'$ jets versus QCD jets. This serves both as an illustration of the power 
of the technique, and as a pedagogical example. 

 The inputs are grids of $25 \times 25$ pixels (i.e the jet-images of Section~\ref{sec:image}), centered around
the reconstructed jet axes, as described in Section~\ref{sec:samples}.  Only the leading jet in each event is used and it is required to have mass $M \in [65, 95]$ GeV.  For image preprocessing, we use trimming to reduce noise, the leading 2 subjets to align the image, and  translation (based on the jet-image energy centroid), reflection, and equalization to uniformize the jet-images.  

After preprocessing, it is observed that the jet-images varied significantly both with the jet transverse momentum ($p_{T}$) and the $\Delta R$ between the leading two subjets ($\Delta R_{jj}$).  Neither of these variations are surprising, since jet $p_{T}$ differences cause wide variations in the energy distribution within the jet and differences in $\Delta R_{jj}$ lead directly to differences in the separation between features in the jet-image.  As a result, the jet-images are binned in both $p_{T}$ and $\Delta R_{jj}$. We use 50 GeV bins for jets with $p_{T} \in [200, 500]$ GeV, and the $\Delta R_{jj}$ binning of each $p_{T}$ bin is ($\Delta R_{jj}<0.4$, $\Delta R_{jj}\in [0.4, 0.6]$, $\Delta R_{jj}\in [0.6, 0.8]$, $\Delta R_{jj}\in [0.8, 1.0]$, $\Delta R_{jj}\in [1.0, 1.2]$, $\Delta R_{jj}>1.2$).  The $\Delta R_{jj}$ binning is chosen such that resolvable differences in subjet position on the order of the subjet radius would not end up in the same bin and smear the jet images, but different optimizations could be explored\footnote{For instance, a $p_{T}$ dependent $\Delta R$ binning may prove beneficial for the very highly boosted cases in which most W jets have small $\Delta R$ separation.}.

After preprocessing both the W and QCD jets, we train a FLD using
these samples as signal and background respectively.  A separate FLD
is trained for each bin of ($p_{T}$, $\Delta R_{jj}$).  It should be noted that equal numbers of W and QCD jets
are used in the training of each FLD, as this was seen to produce 
better results than using different numbers of events in each sample
(or equivalently different priors for each sample).

\begin{figure}[hbt!]
  \centering
  \subfloat[Fisher-Jet \label{subfig:fishjet}]{
        \includegraphics[width=0.455\textwidth]{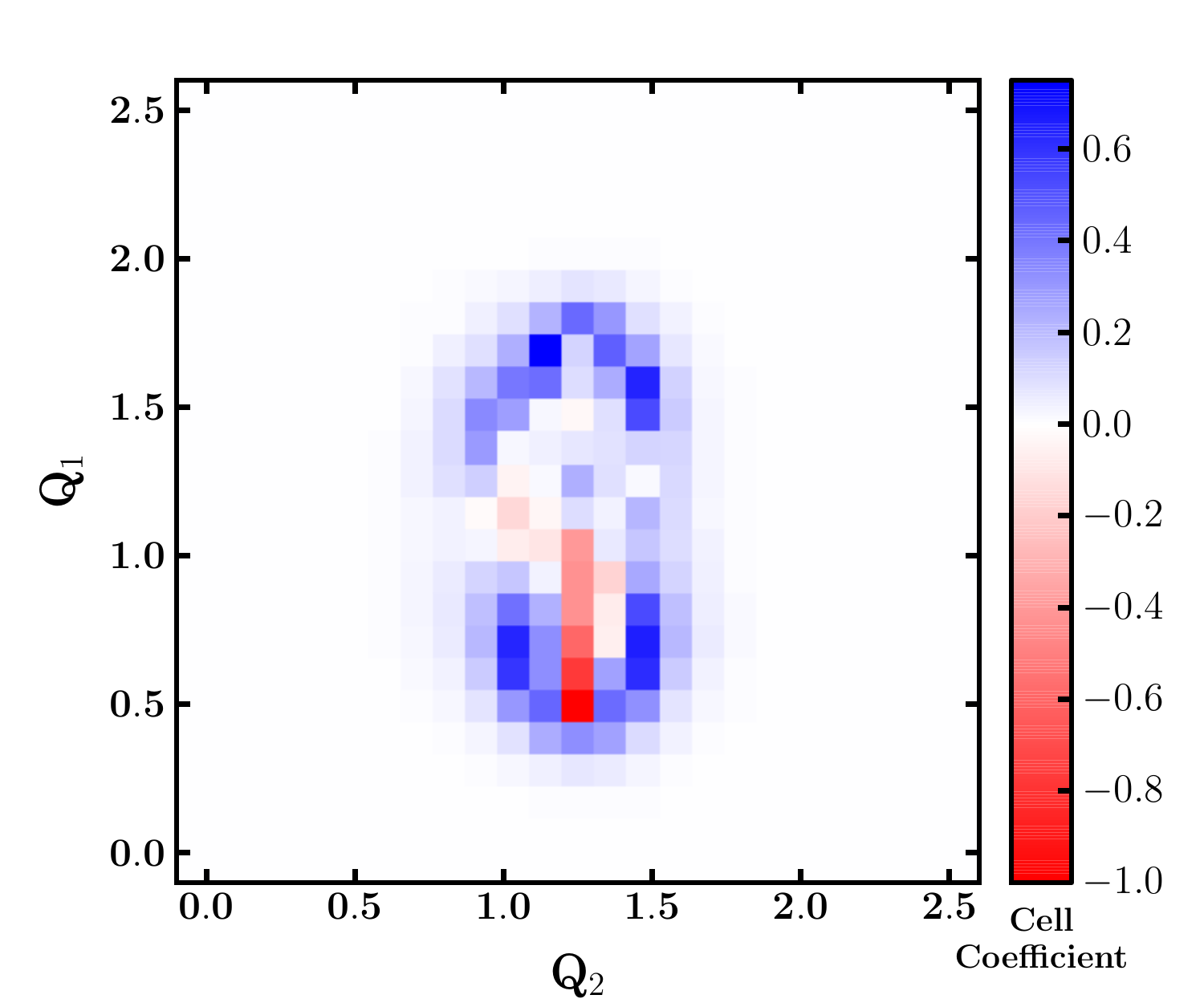}
  }
  \subfloat[Fisher-Jet Discriminant Output\label{subfig:pythiaDisc}]{
    \includegraphics[width=0.53\textwidth]{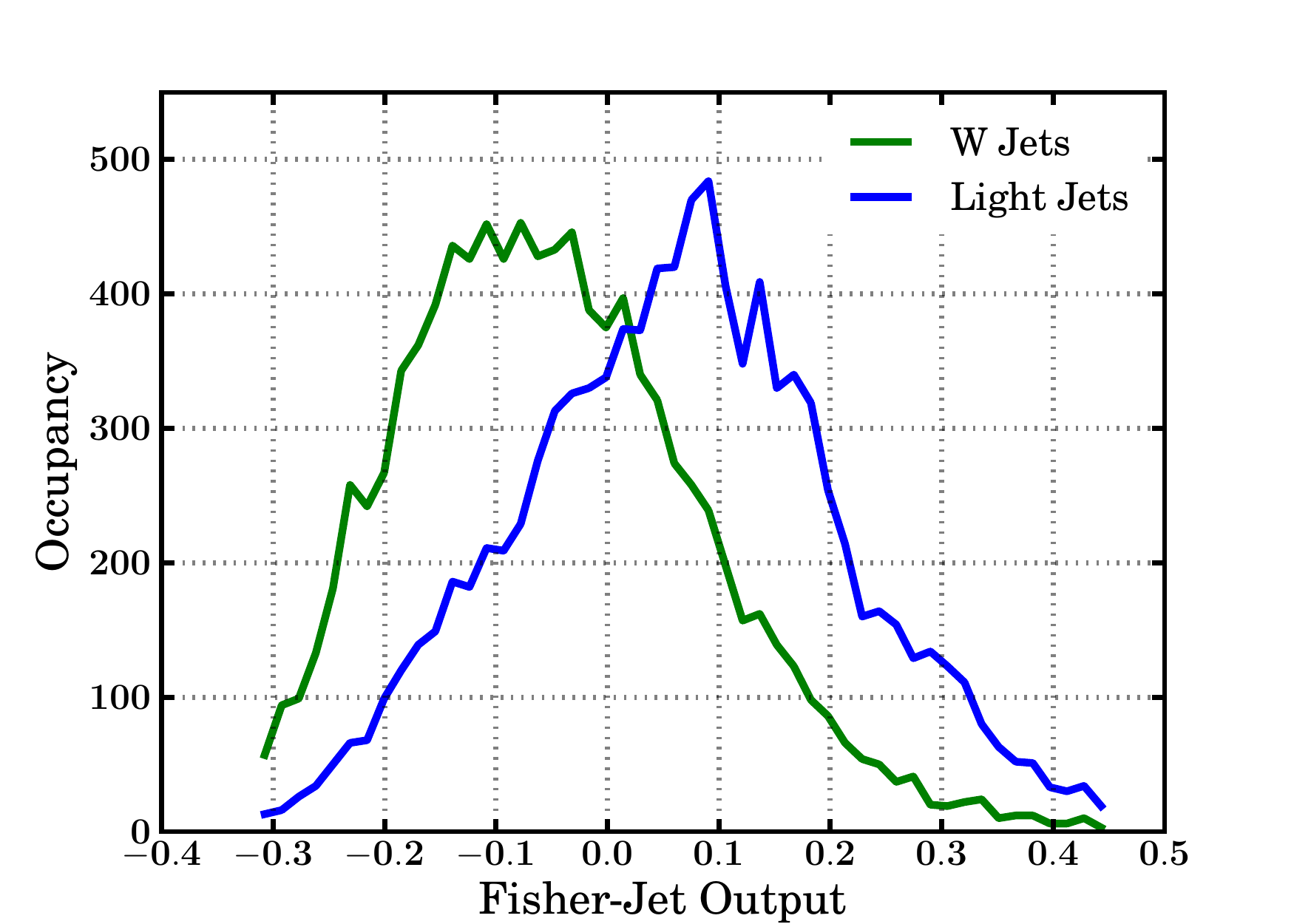}
  }
  \caption{A Fisher's linear discriminant presented as an image (left) and the distributions of the discriminant output when applied to W-jets and Light-jets (right), when the FLD is trained on jets with $p_{T} \in [250, 300]$ GeV, mass $M \in [65, 95]$ GeV, and separation between subjets of $\Delta R \in [0.6, 0.8]$. \label{fig:WLjet} 
  }
\end{figure}

The resulting discriminant Fisher-jet, recast to a $25 \times 25$ image,
 can be seen in Figure~\ref{subfig:fishjet} for a
single bin ($p_{T}\in [250, 300]$ GeV, $\Delta R_{jj}\in$ [0.6, 0.8]),
while the discriminant output when this Fisher-jet is applied to samples of
W and QCD jets can be seen in Figure~\ref{subfig:pythiaDisc}.  
Jet images of several bins of $p_{T}$ and $\Delta R$ can be found in Appendix~\ref{app:fisherimages}.
As noted earlier, the sign of the pixel value in the Fisher-jet indicates
which class the pixel helps identify, as seen in
Figure~\ref{subfig:fishjet} where red pixels indicate importance for W-jets
while blue features imply importance for QCD jets.  We can see in the
figure that the presence of a hard primary subjet located at
($Q_{1}\sim1.5, Q_{2}\sim1.25$) is almost irrelevant for
discrimination, as both classes have hard primary subjets.  However, the
presence of a second hard subjet near ($Q_{1}\sim0.75, Q_{2}\sim1.25$)
is a strong indicator of a W-jet.  Finally, we see that energy
appearing around the two subjets is a strong indication of a QCD jet,
which is essentially telling us that QCD subjets tend to be wider and
have radiation surrounding the subjets.

\begin{figure}[htb!]
  \centering
  \subfloat[\label{subfig:roc6bin}]{
   \includegraphics[width=0.455\textwidth]{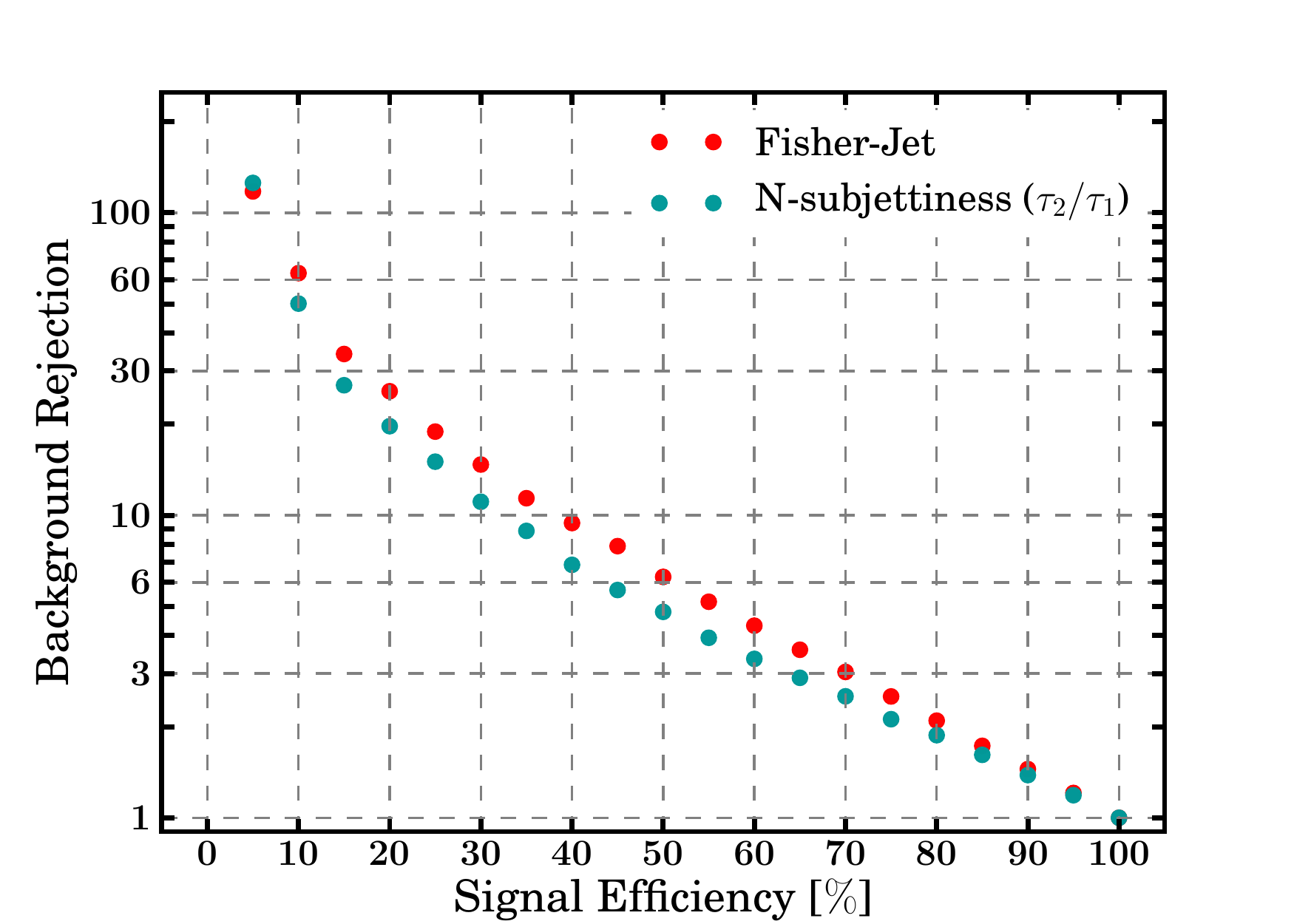}

  }
    \hfill
   \subfloat[\label{subfig:effpt}]{
   \includegraphics[width=0.5\textwidth]{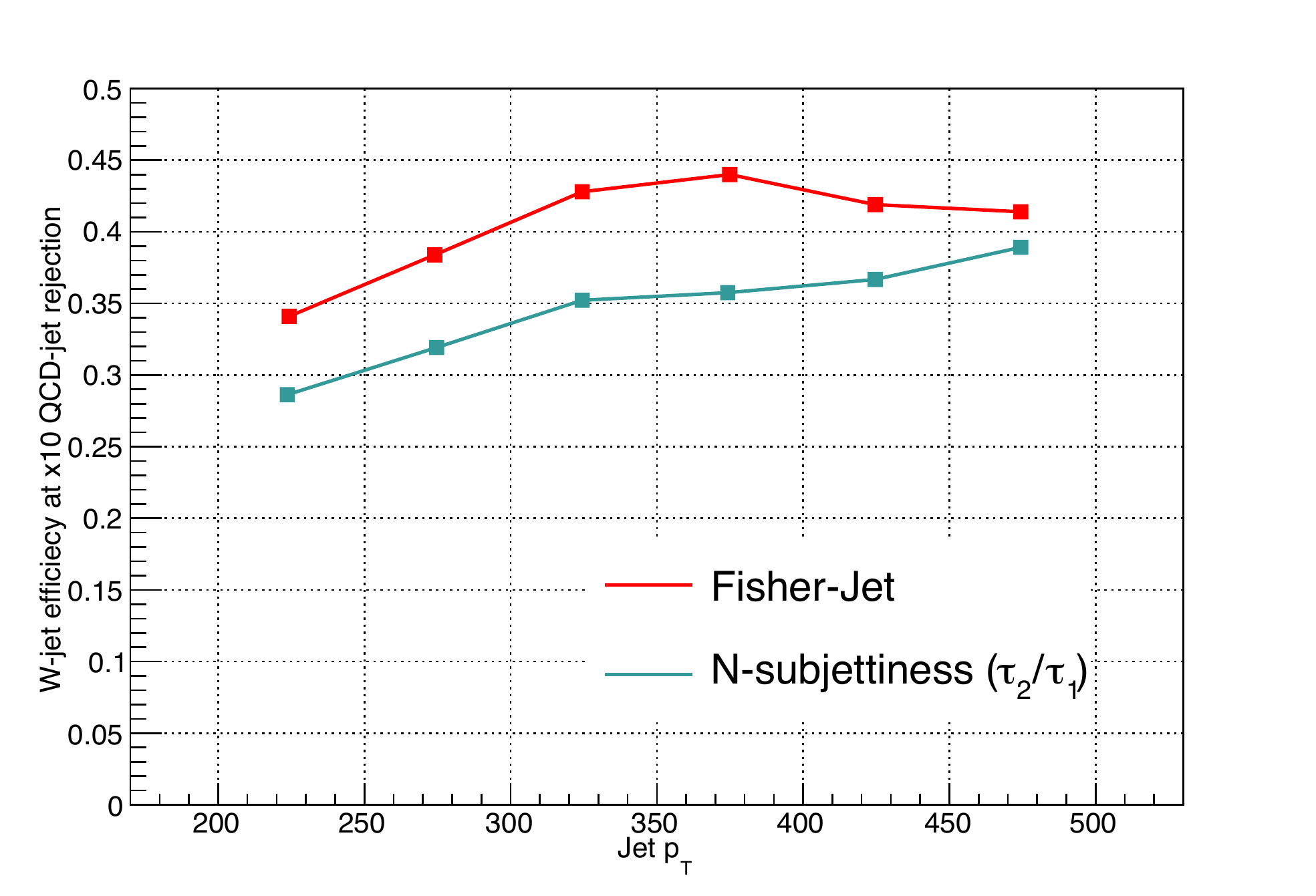}
  }
  \caption{Background rejection  vs signal efficiency curves obtained by training 
    discriminants on samples with $p_{T} \in [250, 300]$ GeV (left), and the W-jet efficiency at fixed $\times10$ QCD jet rejection versus jet $p_{T}$ (right)\label{fig:roc}.
  }
\end{figure}

The background rejection vs. signal efficiency curves for the FLD, computed using the 1-D likelihood ratios of the output distribution of the FLD for W-jets and QCD jets, can be seen in Figure~\ref{subfig:roc6bin}, along with the rejection vs. efficiency curves observed when using N-subjettiness ($\tau_{2}/\tau_{1}$)~\cite{Thaler2011Nsub,Thaler2011Nsub2} computed analogously with the 1-D likelihood ratios.  For the rejection vs. efficiency curve in Figure~\ref{subfig:roc6bin} Fisher-jets are trained on jets satisfying $p_{T} \in [250, 300]$ in 6 bins of $\Delta R_{jj}$, and a combined 1D likelihood ratio distribution is computed by taking the likelihood ratio for each jet computed with respect to appropriate $\Delta R_{jj}$ bin and merging these likelihood ratio values into a single distribution. The N-subjettiness distributions are not binned in $\Delta R_{jj}$ as this did not show any improvements in performance.  Figure~\ref{subfig:effpt} shows the efficiency of W jets at a fixed QCD jet rejection of 10 as a function of jet $p_{T}$ for the FLD (combining the 6 bins of $\Delta R_{jj}$ for each jet $p_{T}$ bin) and for N-subjettiness.  It can be seen that FLD outperforms N-subjettiness for the full range of jet  $p_{T}$ examined.

\begin{figure}[hbt!]
  \centering
  \subfloat[W Jets: FLD output vs $(\tau_2/\tau_1)$\label{subfig:nsubjetiness_w}]{
     \includegraphics[width=0.45\textwidth]{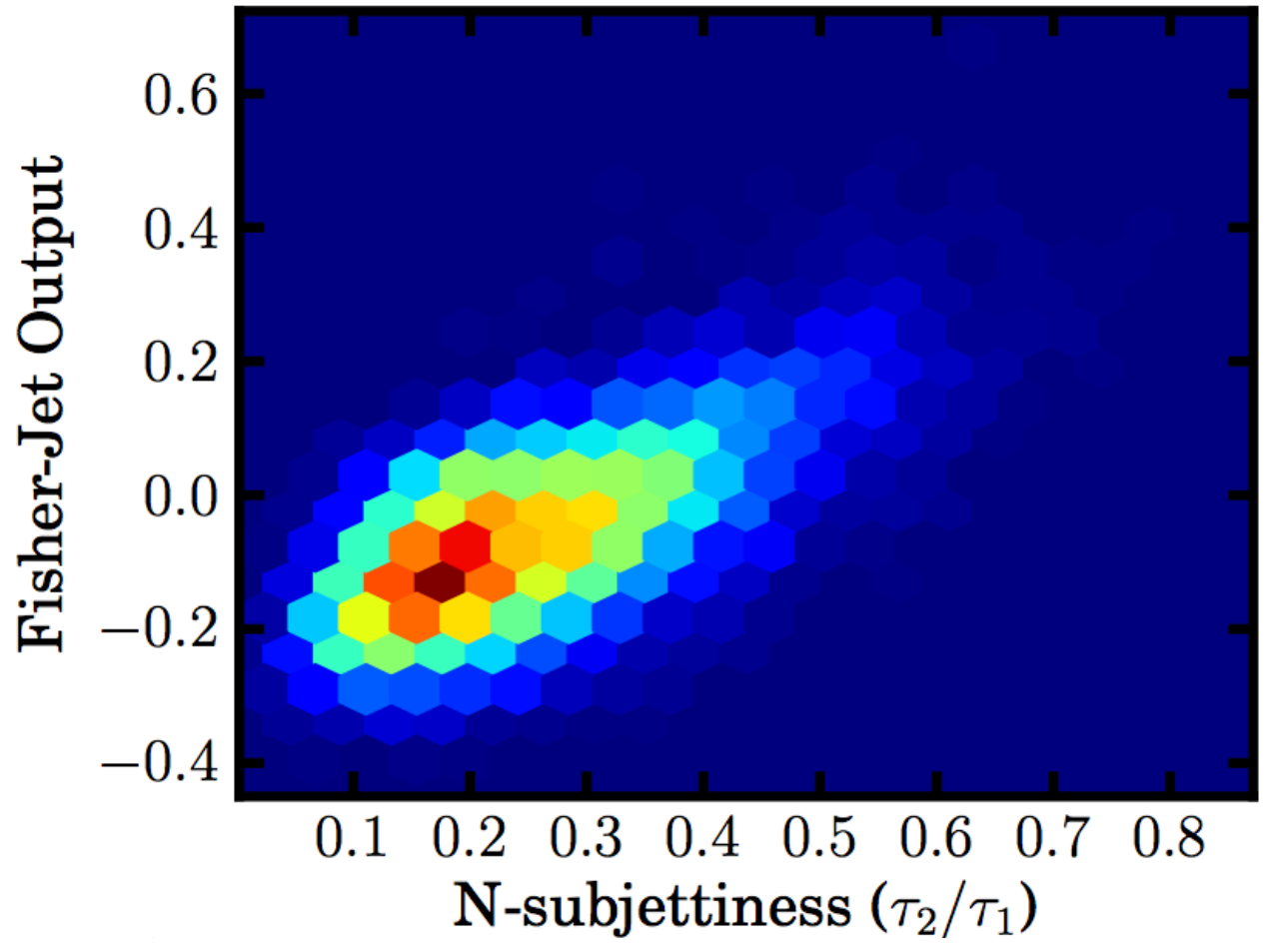}
  }
  \subfloat[QCD Jets: FLD output vs $(\tau_2/\tau_1)$\label{subfig:nsubjetiness_l}]{
    \includegraphics[width=0.45\textwidth]{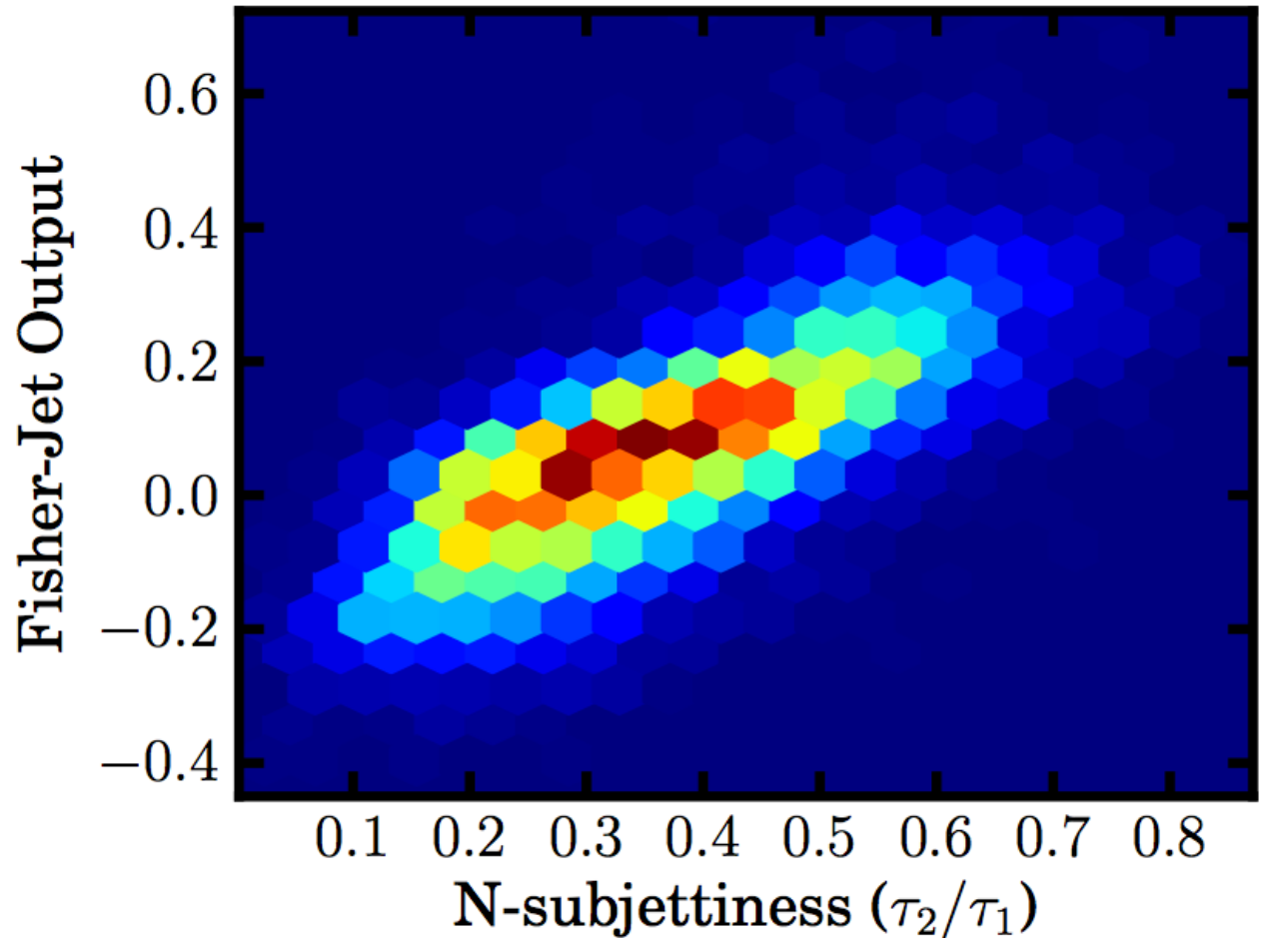}
  }
  \hfill
  \subfloat[W Jets: FLD output vs jet mass\label{subfig:mass_w}]{
    \includegraphics[width=0.45\textwidth]{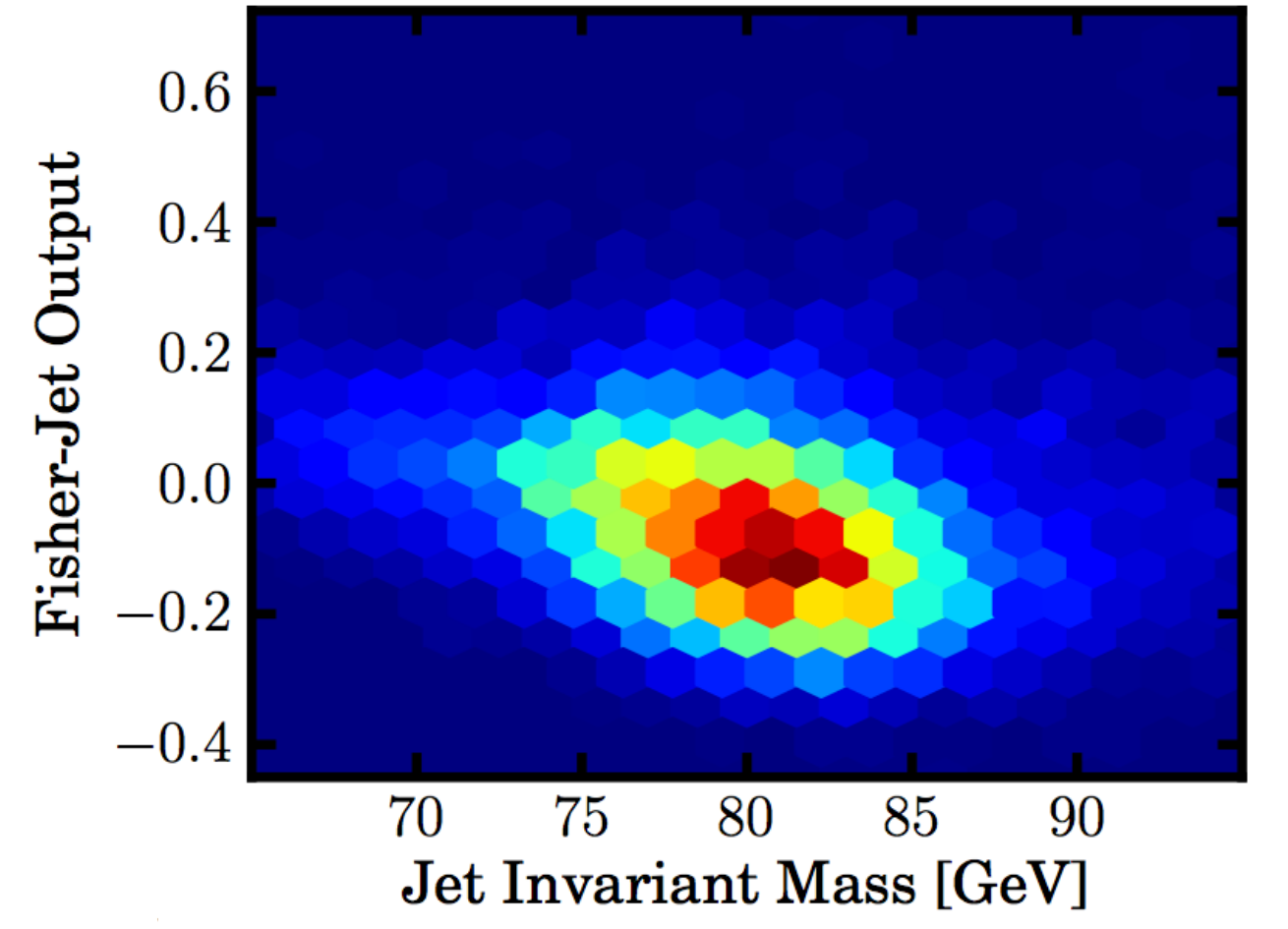}
  }
  \subfloat[QCD Jets: FLD output vs jet mass\label{subfig:mass_l}]{
    \includegraphics[width=0.45\textwidth]{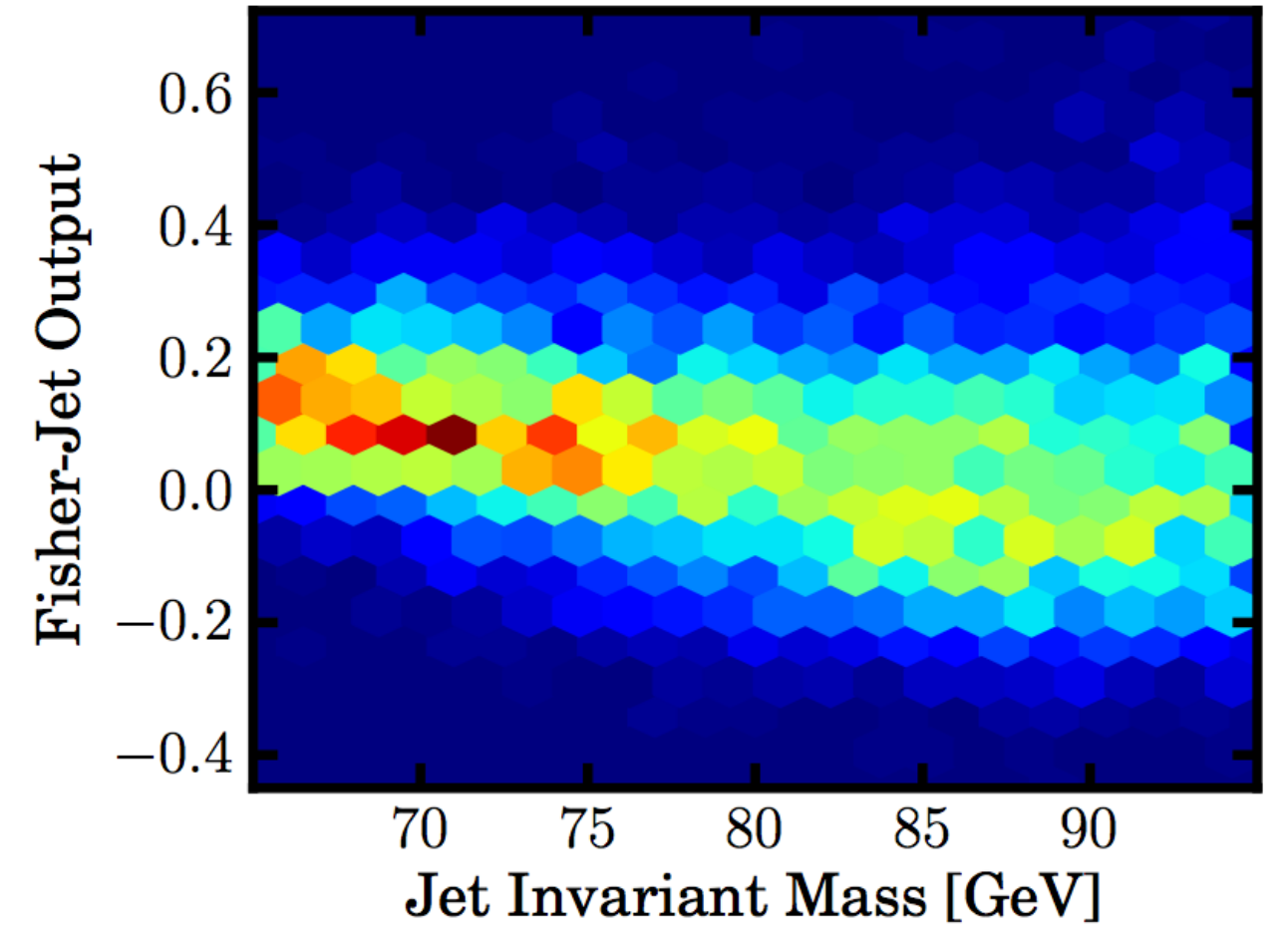}
  }\caption{The Fisher-jet discriminant output for the bin $p_{T} \in [250, 300]$ GeV  and $\Delta R \in [0.6, 0.8]$ plotted against: (\ref{subfig:nsubjetiness_w}) the output of N-subjettiness for W-jets, (\ref{subfig:nsubjetiness_l}) the output of N-subjettiness for QCD jets, (\ref{subfig:mass_w}) the invariant mass of the jet for W-jets, and  (\ref{subfig:mass_l}) the invariant mass of the jet for QCD jets.\label{fig:corr}
}
\end{figure}

It should be noted that the output of FLD and N-subjettiness are
correlated, as shown in Figures~\ref{subfig:nsubjetiness_w} and~\ref{subfig:nsubjetiness_l}
for W and QCD jets respectively, with a correlation coefficient of approximately 0.7 for both W and QCD jets. 
Thus, the Fisher-jet approach is able to combine in a linear way the information
comprising the jet effectively, and capture much of the information of
N-subjettiness and more.  On the other hand, mass, which relies on
quadratic relationships between the inputs, is a simple
quantity which FLD does not reproduce, as shown in
Figures~\ref{subfig:mass_w} and~\ref{subfig:mass_l} for W and QCD jets respectively. 
Since the Fisher-jet output is only slightly correlated with mass, with a correlation 
coefficient of approximately -0.25 for both W and QCD jets indicating a
small degree of anti-correlation, the performance of the classifier
does not change dramatically whether it is applied to a small window
around the $W$ mass, or to a sample without jet mass cuts.

To investigate the effect of pileup, which essentially acts as a source of noise within the jet-image, the Fisher-jets are trained on samples without pileup and subsequently applied to statistically independent samples with pileup\footnote{The reason for such an approach is that the samples without pileup are the best representation of the underlying physics (in the parlance of computer vision, these are the clearest pictures without noise) and are observed to give the best discrimination performance even in samples with noise due to pileup.}. No significant degradation in performance is observed, likely due to the application of trimming. More in depth studies of pileup impact on the Fisher-jet approach are left for future studies.

\begin{SCfigure}
  \centering
  \vspace{-5mm}
  \caption{The Fisher-jet discriminant output, trained on a sample of
    \pythia $W$ and QCD jets applied to \pythia and \herwig
    samples.\label{fig:generator}
    }
    \includegraphics[width=0.5\textwidth]{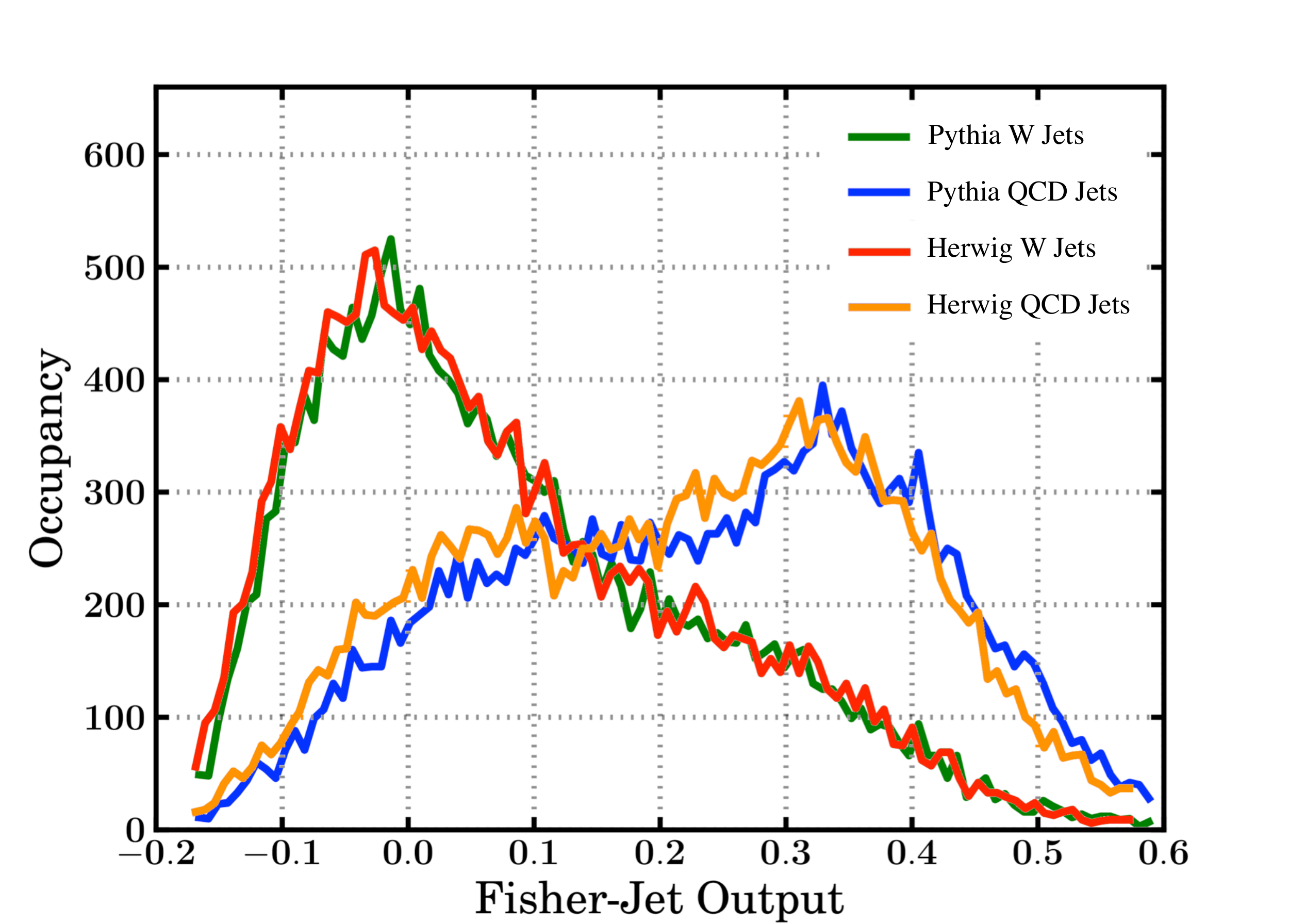}
\end{SCfigure}

To check the generator dependence of the classifier, a second sample
of $W$ and QCD jets are generated using \herwig which implements an
independent description of the hard sub-process and subsequent
showering. The Fisher-jet trained on the \pythia samples
is then applied to the \herwig samples. The Fisher-jet discriminant output distributions
for both W-jets and QCD jets from both \pythia and \herwig are shown
Figure~\ref{fig:generator} for jets with $p_{T} \in [200, 250]$ GeV and  $\Delta R \in [0.6, 0.8]$, and are seen to be extremely similar between both generators. 

\clearpage

\section{Conclusion}
\label{sec:conclusion}
We present a generic algorithm to extract discriminating information between different classes of jets using parallels to techniques used in the field of computer vision\footnote{The software used in this analysis is available at  \href{https://github.com/makagan/JetImages}{https://github.com/makagan/JetImages}.}.  The algorithm uses a representation of jets as images, applies preprocessing techniques to construct a consistent set of jet-images, and applies a Fisher linear discriminant, which has been trained on a collection of example jets, to process the jet-images.  The application of the Fisher discriminant projects different classes of jets to opposite ends of the discriminant output spectrum, thereby allowing for separation of the classes.  The resulting algorithm is a linear function of the inputs and is easy to understand, with little prior  knowledge needed in the tuning of the discriminant and does not rely on an analytic description of the system studied.  This method is seen to be competitive with N-subjettiness when separating hadronically decaying boosted $W$ bosons from QCD jets.

Importantly, the resulting algorithm can be used as a tool to
visually discover differences in samples with no complete analytical
description.  The jet-images and the Fisher-jet can be trivially visualized,
allowing one to study the important features of different jet classes and
to gain insight into the features which are most discriminating between different classes.
In this way, the method provides a starting place to provide simple answers to
high dimensional problems.

This computer vision inspired approach stands out in its flexibility and its
ease of inspection.  While this study focuses on $W$ vs QCD jets,  initial studies
have show that the method can be easily applied to the $H\rightarrow
b\bar{b}$ system and provide powerful discrimination.  The algorithm could be adapted  for use in other problems of jet discrimination, such as identifying hadronically decaying tops. 
More generally, many problems in experimental physics contain a
correspondence to the imaging problem discussed here, such as
categorizing the ring images produced in Super-K~\cite{SuperKDesign}, where 
our methods are likely directly applicable.  
In addition, since a consistent set of jet-images is produced by the pre-processing steps,  non-linear classifiers could easily be used for jet-image processing to improve the discrimination performance. 

Jet-images provide a connection between the fields of particle physics and computer vision, and is a departure from the ways that experimental high energy physics currently uses machine learning algorithms.
However, this is merely a first step, while many of the techniques of the vast field of computer vision
have yet to be explored in the context of particle physics and may prove powerful within this domain.

\section*{Acknowledgements}
We would like to thank Jon Butterworth, Matthew Schwartz, Peter Skands, and Jesse Thaler for their helpful feedback.  We would also like to thank Su Dong and the SLAC ATLAS group for their insightful discussions on the topic. This work is supported by the US Department of Energy (DOE) Early Career Research Program and grant DE-AC02-76SF00515.

\clearpage
\begin{appendices}
\section{Fisher Jet Images}
\label{app:fisherimages}

\begin{figure}[hbt!]
  \centering
        \includegraphics[width=0.95\textwidth]{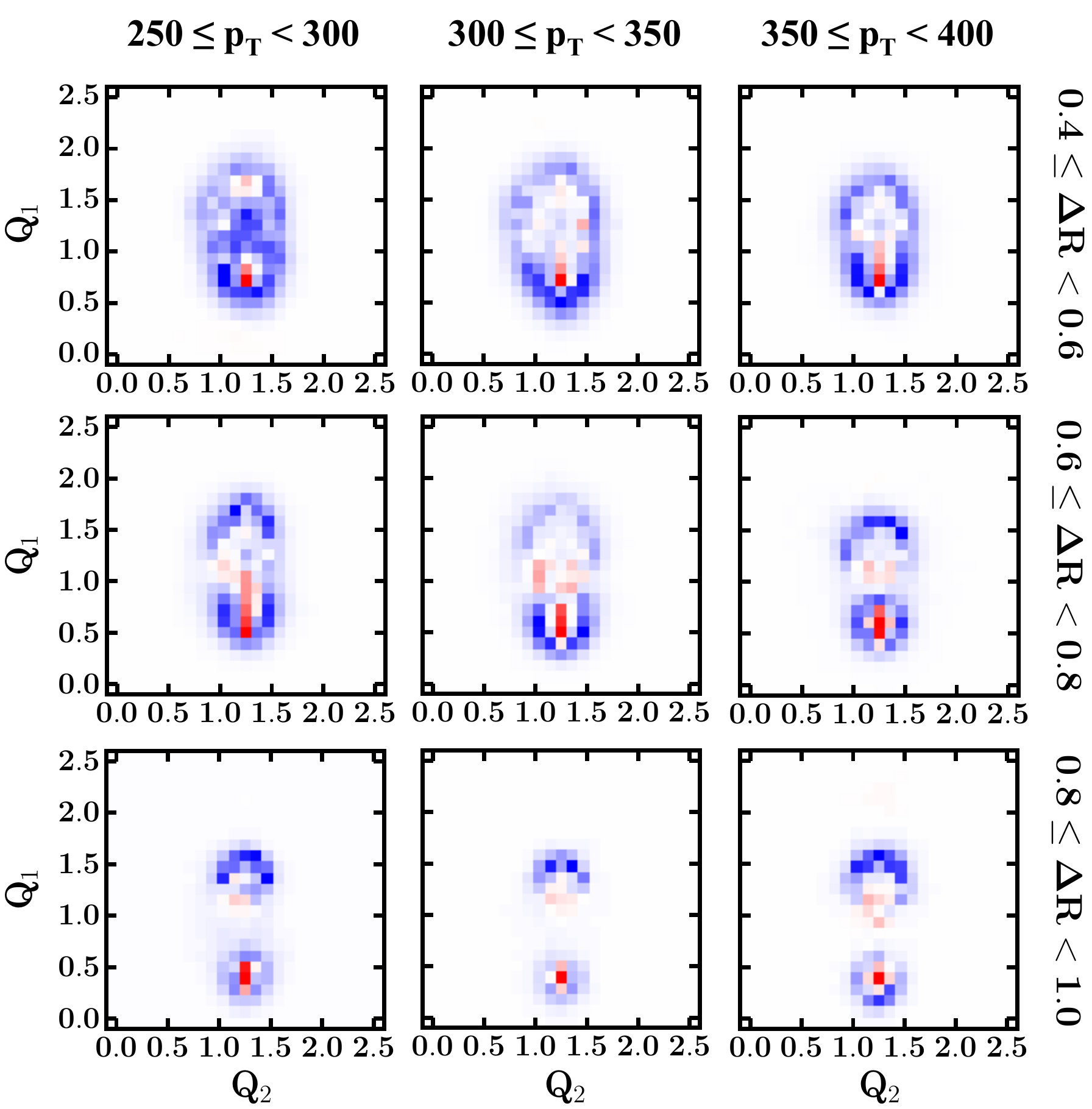}
  \caption{Fisher linear discriminants presented as jet-images, or Fisher-jets, for several $p_{T}$ and $\Delta R_{jj}$ bins for jets with mass $M \in [65, 95]$ GeV. \label{fig:fisherjetimages} 
  }
\end{figure} 
\end{appendices}
\newpage
\bibliography{JetImages}

\end{document}